\documentclass[twocolumn,aps,longbibliography]{revtex4-1}

\usepackage{amsmath,amssymb,graphicx,psfrag,color}

\begin{document}
\title{The real and imaginary parts of a weak value appearing as back-actions via a post-selection}
\author{Kazuhiro Yokota}
\affiliation{Institute of Quantum Science, Nihon University, Chiyoda-ku, Tokyo 101-0062, Japan}

\begin{abstract}
In a weak measurement the real and imaginary parts of a weak value participate in the shifts of the complementary variables of a pointer.
While the real part represents the value of an observable in the limit of zero measurement strength, the imaginary one is regarded as the back-action due to the measurement with a post-selection, which has an influence on the post-selection probability.
In this paper I give a case in which a real part could also appear as such a back-action in a post-selection probability on an equal footing with an imaginary one.
It is also shown that both of the real and imaginary parts can be inferred by observing the probability in practice, which has an advantage that an additional system of a pointer is not needed.
\end{abstract}

\maketitle

\section{Introduction}
Recently a weak value has attracted attention in both of the foundation and the application of quantum mechanics.
The value was originally introduced as a result of a weak measurement \cite{W1}, which gives us a value of an observable, $\hat{O}$, as an ensemble average without disturbing the measured system.
When the system is initially in $|\psi\rangle$ (pre-selection), and is finally found in $|\phi\rangle$ (post-selection), a weak measurement performed between the pre-post-selection shows the weak value of $\hat{O}$ as follows,
\begin{eqnarray}
\langle\hat{O}\rangle_{\bf w}=\langle\phi|\hat{O}|\psi\rangle /\langle\phi|\psi\rangle.  \label{eq:wv}
\end{eqnarray}

Although a pre-post-selection has been often seen as in quantum information processing, the concept of a weal value was inspired by a time symmetric description of quantum mechanics \cite{ABL, time_s}.
Actually a weak measurement has offered a new approach to the foundation of quantum mechanics, for example, in the Leggett-Garg inequality \cite{W_LG1, W_LG2, W_LG3}, the contextuality \cite{W_context1, W_context2, W_context3, W_context4, W_context5} and so on, since we can access a quantum system without disturbance.
In particular a weak measurement has been performed experimentally for observation of a quantum paradox \cite{time_s, W_para1, W_Hardy, W_Hardy1, W_Hardy2, W_para2, W_para3, W_para4, W_para5, W_para6, W_para7, W_para8}.
It has been also shown that a weak value itself plays an important role in a quantum phenomenon irrespective of a weak measurement \cite{W_ph1, W_ph2, W_ph3, W_ph4}.

The large shift of a pointer could be produced in a weak measurement by choosing a pre-post-selection on purpose as $\langle\phi|\psi\rangle\rightarrow 0$ in equation (\ref{eq:wv}). 
Using such an amplification effect, an application for precision measurement has been actually demonstrated \cite{W_amp1, W_amp2, W_amp3, W_amp_t1, W_amp_t2, W_amp_t3, W_amp_t4, W_amp_t5}.
As another application, a weak measurement has been also expected in sensing science \cite{WM_sense1, WM_sense2} beyond the fundamental issue of direct measurement of a quantum state \cite{WM_dir1, WM_dir2}.

According to the definition in equation (\ref{eq:wv}), a weak value is generally a complex number.
In a weak measurement the real and imaginary parts of a weak value appear as the shifts in the complementary variables of a pointer:
For example, the position of the pointer, $x$, shifts in response to the real part, while the imaginary one participates in the shift of the momentum, $p$ ($[x,p]=i\hbar$).

Despite of such a similarity the interpretation of each parts has been actively discussed:
While the real part has been naively considered as the value of the measured observable in the limit of zero disturbance, the imaginary one seems to represent the disturbance (back-action) due to the measurement with a postselection \cite{W_im1, WV1, WV2}.

Roughly speaking, the real part of a weak value has mostly played a significant role in a fundamental issue as in a quantum paradox so far.
On the other hand, an imaginary part has been found to be practically useful in the application of an amplification for precision measurement \cite{W_im4, W_im3}, while recently it was reported that an appearance of an imaginary part can be associated with contextuality \cite{W_context5}.

In this paper I show a case that the real part of a weak value also could have an influence on the post-selection probability as a back-action in the same manner as the imaginary one.
While I expect such a suggestion to open up a new approach for deeper understanding of a weak value as trying to treat both of the parts on an equal footing, I would like to discuss an application for estimating a weak value experimentally on this occasion:
Without preparing an additional system of a measurement apparatus, both of the real and imaginary parts can be inferred by observing the post-selection probability affected by their back-actions.

\section{The real and imaginary parts of a weak value in a weak measurement}
\label{sec:WM}
First I would like to review a weak measurement by using von Neumann measurement model as originally proposed in \cite{W1}, and introduce the real and imaginary parts of a weak value.

A measurement apparatus is prepared as in the state of $|\Psi(Q)\rangle$, where $Q$ represents the position of a pointer which gives us a result of measurement.
I assume the distribution of the position is provided by a Gaussian function as follows,
\begin{eqnarray}
\langle Q|\Psi(Q)\rangle = {\rm exp}\bigg(-\frac{Q^2}{2\sigma^2}\bigg)
\end{eqnarray}
up to normalization, where $\sigma^2$ presents the variance.
To observe $\hat{O}$ on a quantum system, $|\psi\rangle$, the system is interacted with the pointer by the interacting Hamiltonian as follows,
\begin{eqnarray}
\hat{H}=g(t)\hat{O}\hat{P},   \label{eq:H}
\end{eqnarray}
with the momentum of the pointer, $\hat{P}$; the coupling function, $g(t)$, satisfying $\int dtg(t)=G$ for the duration of the interaction.
As a result the system is correlated with the pointer as follows,
\begin{eqnarray}
|\psi\rangle\otimes|\Psi(Q)\rangle\rightarrow \sum_k\langle o_k|\psi\rangle|o_k\rangle\otimes|\Psi(Q-Go_k)\rangle,
\end{eqnarray}
where I have assumed $\hat{O}$ has the discrete spectrum of an eigenvalue, $o_k$, with an eigenstate of $|o_k\rangle$.

The coupling strength can be controlled by $G$, which corresponds to the measurement strength.
Actually, when $G$ is large enough, we can certainly discriminate which eigenstate the system is in by reading the shift of the pointer in response to the eigenvalue, $\Delta Q=Go_k$.
Then the system is utterly disturbed by the measurement, as the system results in one of the eigenstates, $|o_k\rangle$.
As such an observation of the eigenstate appears with a probability, $|\langle o_k|\psi\rangle|^2$, the average of the pointer shift gives us the ensemble average, $\bar{O}=\langle\psi|\hat{O}|\psi\rangle$.
In this case the measurement model well represents a conventional measurement process, which is so-called a strong measurement.

On the other hand, given $\sigma$, a weak measurement is achieved in $G\rightarrow 0$.
The system is not disturbed due to the almost no correlation, which seems a failure of measurement.
Even in this case, however, the pointer contains a piece of information on the system.
Actually, if the ensemble average is taken, the probability distribution of the pointer position is given by $Prob(Q) \sim |\Psi(Q-G\bar{O})|^2$.
Unlike a strong measurement, the ensemble average, $\bar{O}$, is obtained without disturbance on the system.

Furthermore, when the system is finally post-selected in $|\phi\rangle$, the shift of the pointer is given by the weak value as follows, 
\begin{eqnarray}
& &\langle\phi|e^{-\frac{i}{\hbar}G\hat{O}\hat{P}}|\psi\rangle\otimes|\Psi(Q)\rangle \nonumber \\
&\sim& \langle\phi|(1-\frac{i}{\hbar}G\hat{O}\hat{P})|\psi\rangle \otimes |\Psi(Q)\rangle \nonumber \\
&\sim& \langle\phi|\psi\rangle e^{-\frac{i}{\hbar}G\langle\hat{O}\rangle_{\bf w}\hat{P}}|\Psi(Q)\rangle.  \label{eq:WM}
\end{eqnarray}

As shown in equation (\ref{eq:wv}), the weak value is generally a complex number;
The real and imaginary parts of the weak value appear in the shifts of the position, $\Delta Q=G{\rm Re}\langle\hat{O}\rangle_{\bf w}$, and the momentum, $\Delta P=2G{\rm Im}\langle\hat{O}\rangle_{\bf w}/\sigma^2$, respectively \cite{W_com}.
Then, as in the case of a conventional strong measurement, the real part is simply obtained by the shift of the position, $\Delta Q$, normalized by the measurement strength, $G$, i.e. $\Delta Q/G$.
On the imaginary part, the momentum shift contains the variance of the pointer $\sigma^2$ unlike the real one.

Actually, without reference to von Neumann measurement model, it has been known that the shift corresponding to the imaginary part depends on the details of the pointer.
As a result it gives rise to the different interpretations of the real and imaginary parts of a weak value.
While the real part can be regarded as the conditioned average of $\hat{O}$ in the limit of zero disturbance, the imaginary one can not be naively associated with the measurement of $\hat{O}$;
Rather the imaginary part is interpreted as disturbance (back-action) \cite{W_im1}, which could be confirmed in the success probability of the post-selection as discussed in the next section.

\section{A weak value appearing as a back-action via a post-selection}
\label{sec:main}
In \cite{W_im1} it was clarified what the imaginary part of a weak value represents in a weak measurement:
An imaginary part provides information about how the initial state is disturbed by the observable operator, which could be confirmed in the change of post-selection probability.

To see this, I re-describe the Hamiltonian of equation (\ref{eq:H}) with $\int dtg(t)\hat{P}/\hbar=G\hat{P}/\hbar\equiv\theta$ as in \cite{W_im2} without referring to a pointer.
As the momentum, $P$, is a constant of motion under the Hamiltonian, I have regarded $\theta$ as just a parameter.
Then the post-selection probability is given as follows,
\begin{eqnarray}
Prob(\phi)&=&|\langle\phi|e^{-i\theta\hat{O}}|\psi\rangle|^2 \nonumber \\
&\sim& |\langle\phi|\psi\rangle|^2(1+2\theta{\rm Im}\langle\hat{O}\rangle_{\bf w}),   \label{eq:post}
\end{eqnarray}
in $\theta <<1$ corresponding to weak measurement.
I refer to the change of the post-selection probability due to the weak value as `back-action.'
In fact I will show that such a back-action in the post-selection probability could be also come into by not only the imaginary part of a weak value but also the real part.

\begin{figure}
 \begin{center}
	\includegraphics[bb=0 0 481.400204 223.181550, width=0.5\textwidth]{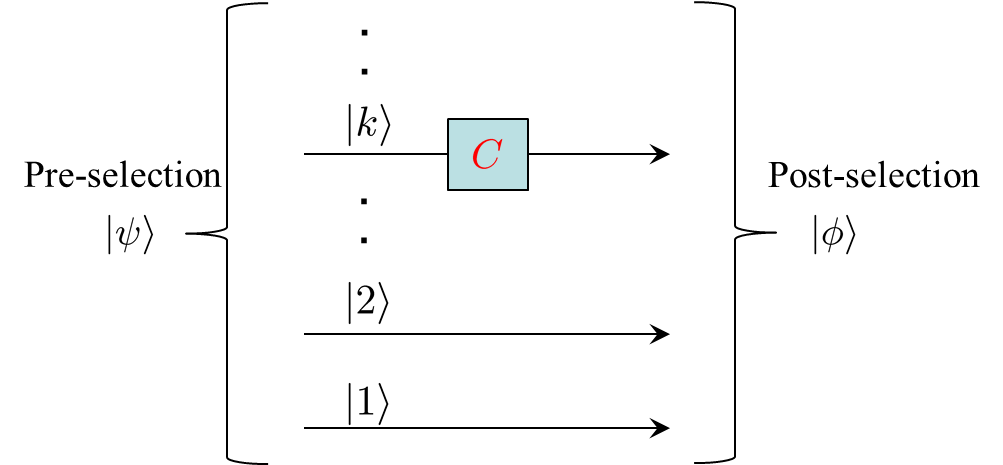}
 \end{center}
	\caption{A photon takes the superposition of the path state, $|k\rangle$, which is pre-post-selected in $|\psi\rangle$ and $|\phi\rangle$.
}
\label{fig:path}
\end{figure}

When deriving equation (\ref{eq:post}), I have payed attention to only the measured system.
Actually the back-action given by a weak value does not need to rely on the context of a weak measurement any longer.
To clarify this point, it will be helpful to treat a specific case of a photon.

Suppose that a photon can take paths, $\{|k\rangle\}$, and the initial state is in the superposition as follows,
\begin{eqnarray}
|\psi\rangle = \sum_{k}\langle k|\psi\rangle|k\rangle.
\end{eqnarray}
Additionally the path state is finally post-selected in $|\phi\rangle$, and I would like to discuss one of the weak values of the projectors, $|k\rangle\langle k|$.
As shown in figure \ref{fig:path}, an optical component is placed on the path, $|k\rangle$, by which the corresponding term is multiplied by a c-number, $C$: $|k\rangle\rightarrow C|k\rangle$.

In this case the back-action in the post-selection probability in equation (\ref{eq:post}) can be imitated by setting the phase shifter, $C=e^{-i\theta}$, on the path of $|k\rangle$: the probability changes as follows,
\begin{eqnarray}
Prob(\phi)&=&|\langle\phi|e^{-i\theta}|k\rangle\langle k|\psi\rangle|^2 \nonumber  \\
&\sim& |\langle\phi|\psi\rangle|^2(1+2\theta{\rm Im}\langle|k\rangle\langle k|\rangle_{\bf w}),  \label{eq:post_im}
\end{eqnarray}
in $\theta << 1$.

On the other hand, when the optical component is an attenuator with the transmittance, $T=1-R$, i.e. $C=\sqrt{T}\equiv e^{-\alpha}$ ($\alpha \ge 0$), the post-selection probability is given as follows,
\begin{eqnarray}
Prob(\phi)&=&|\langle\phi|e^{-\alpha}|k\rangle\langle k|\psi\rangle|^2  \label{eq:post_re_ex} \\
&\sim& |\langle\phi|\psi\rangle|^2(1-2\alpha{\rm Re}\langle|k\rangle\langle k|\rangle_{\bf w}) \label{eq:post_re} \\
&\sim& |\langle\phi|\psi\rangle|^2(1-R{\rm Re}\langle|k\rangle\langle k|\rangle_{\bf w}),  \label{eq:post_re_R}
\end{eqnarray}
where I have assumed $\alpha <<1$ ($R << 1$).

Generally when a c-number, $C_k$, is applied to the path, $|k\rangle$, the post-selection probability is given as follows,
\begin{eqnarray}
Prob(\phi)=|\langle\phi|\psi\rangle|^2\bigg|\sum_k C_k\langle|k\rangle\langle k|\rangle_{\bf w}\bigg|^2 ;   \label{eq:postprob}
\end{eqnarray}
If $C_k$ satisfies the condition of a weak disturbance like $\theta, \alpha <<1$, a weak value of each path explicitly appears as in equations (\ref{eq:post_im}) and (\ref{eq:post_re}).

In \cite{W_negS} it was already shown that a real weak value is useful in estimating how the post-selection probability is change by dissipations which are placed between the pre-post-selection:
A negative real weak value well represents a negation of a dissipation against a positive one.
Especially, when $\langle |k\rangle\langle k|\rangle _{\bf w}=1$ in equation (\ref{eq:post_re_ex}) without the approximation, the post-selection probability decreases as if a photon has passed the attenuator with certainty, i.e. $Prob(\phi)=T|\langle\phi|\psi\rangle|^2$.
In addition the same attenuator is added on another path of $\langle|k'\rangle\langle k'|\rangle_{\bf w}=-1$, the probability restores to $|\langle\phi|\psi\rangle|^2$, according to equation (\ref{eq:postprob}).

In this discussion of \cite{W_negS}, the negative weak value of $-1$ by itself does not make sense, since the negative value cannot be related to `negation' without the positive value.
In other words the negative weak value of $-1$ is meaningful in the equation, $1-1=0$.  

However, under the weak condition of $R<<1$ in equation (\ref{eq:post_re_R}), the negative weak value itself can be a counterpart of the positive weak value:
While the positive weak value of $1$ gives the post-selection probability of $Prob(\phi)=(1-R)|\langle\phi|\psi\rangle|^2$, the negative value of $-1$ gives the probability, $Prob(\phi)=(1+R)|\langle\phi|\psi\rangle |^2$.
Although such a symmetric relation of the positive weak value $1$ and the negative one $-1$ was also confirmed as linear-polarization shift in \cite{W_negW}, I have shown that the symmetric relation is also found in the post-selection probability without another system of polarization.
Actually the supplementary result of Figure.5(b) in \cite{W_negS} implies such a symmetric relation when $T$ is large (i.e. $R<<1$).

Another point to note is when the component on the path of $|k\rangle$ provides a c-number, $C=e^{-\alpha}e^{-i\theta}$.
In this case the probability is simply given by
\begin{eqnarray}
Prob(\phi) \sim |\langle\phi |\psi\rangle |^2(1+2\theta {\rm Im}\langle |k\rangle\langle k|\rangle _{\bf w} \nonumber \\
\ \ \ \ \ \ \ \ \ \ \ \ \ \ \ \ \ \ \ \ \ \ \ \ \ \ \ \ -2\alpha {\rm Re}\langle |k\rangle\langle k|\rangle _{\bf w}),
\end{eqnarray}
with $\theta,\alpha$ $<<1$.
Both of the real and imaginary parts of a weak value appear in an equal footing.
Clearly such equivalent contributions to the post-selection probability by both of the parts are also found in another situation, for example, when components are set on some paths (not only on one path). 
As shown in equation (\ref{eq:postprob}), the c-number on each path is weighted by the corresponding weak value.
Under the weak condition ($\theta,\alpha <<1$), the real and imaginary parts of a c-number are weighted by the real and imaginary parts of the corresponding weak value respectively.

\section{An application of the back-action by a weak value}
\label{sec:app}
\begin{figure}
 \begin{center}
	\includegraphics[bb=0 0 608.923885 827.896513, width=0.4\textwidth]{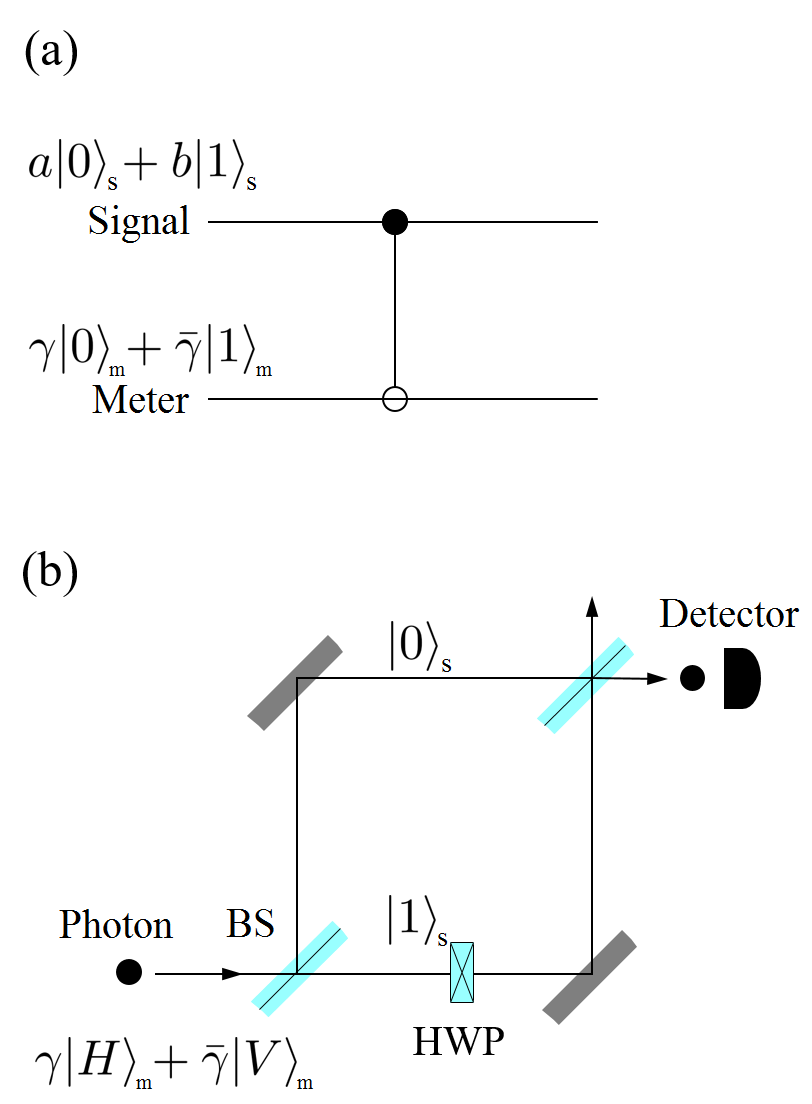}
 \end{center}
	\caption{(a)A CNOT operation for a weak measurement on a qubit. The signal qubit is correlated with the meter qubit, whose strength is controlled by the input state of the meter, namely $\gamma$ and $\bar\gamma$.
(b)A weak measurement on the path state of a photon, $\langle|1\rangle_s\langle 1|\rangle_{\bf w}$ and $1-\langle|1\rangle_s\langle 1|\rangle_{\bf w}=\langle|0\rangle_s\langle 0|\rangle_{\bf w}$.
The pre-post-selection on the path state is accomplished by the interferometer composed of the beam splitters (BS) and the detection of one of the output ports.
The horizontal (vertical) polarization, $|H\rangle_m$ ($|V\rangle_m$) corresponds to $|0\rangle_m$ ($|1\rangle_m$) of the meter qubit.
The CNOT operation is achieved by the half wave plate (HWP), by which the polarization is changed as $|H\rangle_m\leftrightarrow|V\rangle_m$.
}
\label{fig:WM}
\end{figure}
\begin{figure}
 \begin{center}
	\includegraphics[bb=0 0 740.038030 961.049387, width=0.5\textwidth]{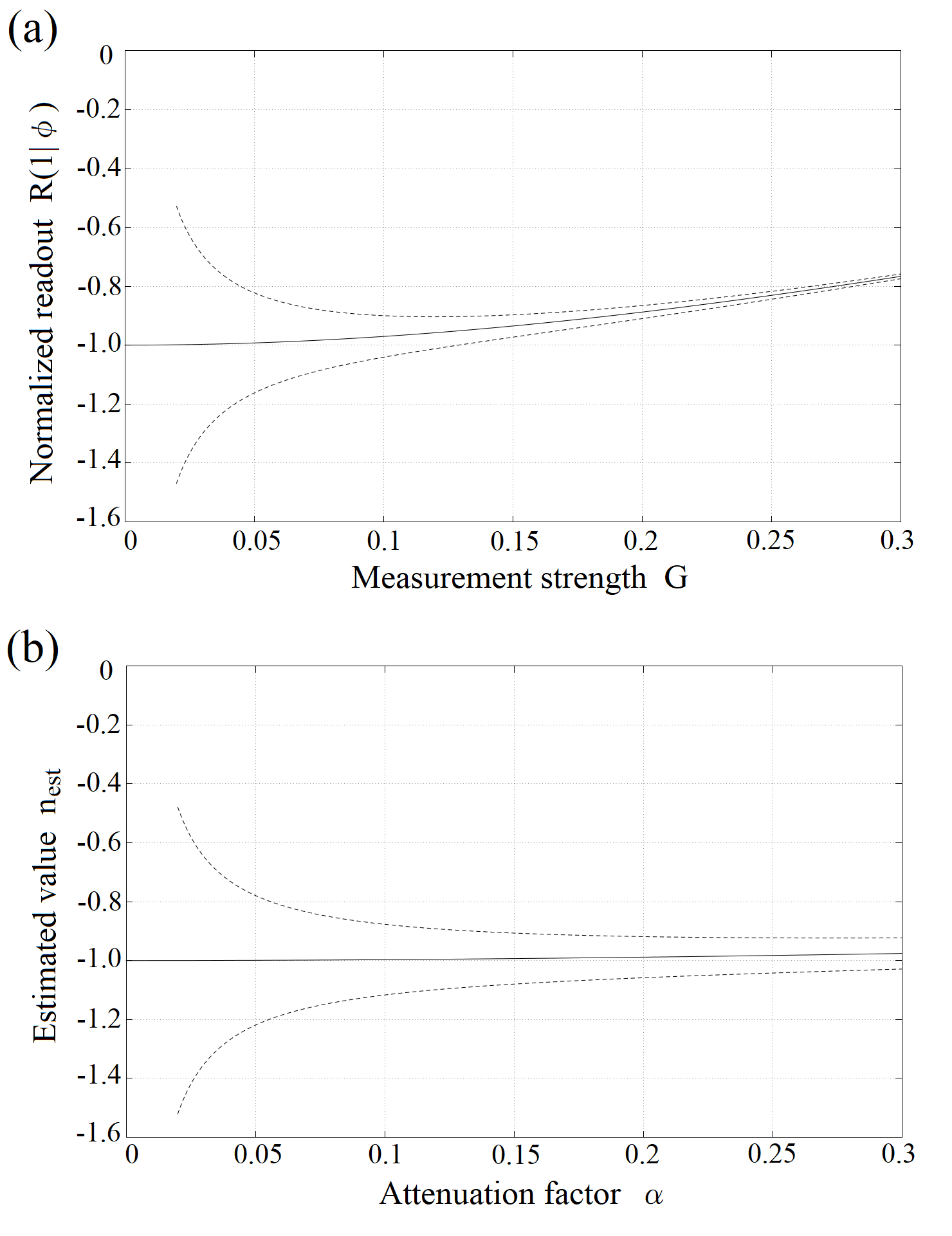}
 \end{center}
	\caption{(a)The normalized readout. (b)The estimated value from the post-selection probability.
In the limit of the measurement strength $G$ $\rightarrow$ $0$ and the attenuation factor $\alpha$ $\rightarrow$ $0$, they show the real part of a weak value, $-1$.
The solid lines correspond to the ideal values of equation (\ref{eq:norm}) and (\ref{eq:est}), while the dashed lines represent the errors stemming from the Poisson statistics of the photon counts: the photon counts of the post-selection, $|\langle\phi|\psi\rangle|^2$, is assumed to be $10000$, based on which the other counts are considered to calculate the probabilities, $Prob(1|\phi)$ and $Prob(\phi)$.
}
\label{fig:est}
\end{figure}

I have shown that both of the real and imaginary parts of a weak value could contribute to the post-selection probability.
In this section I would like to discuss an application of this result, that is, whether it is practical to estimate both of the parts by observing the post-selection probability in equations (\ref{eq:post_im}) and (\ref{eq:post_re}).

For experimentally observing a weak value, it is straightforward to use a weak measurement as reading the pointer shift in equation (\ref{eq:WM}).
There are also experimental approaches to measure the real and imaginary weak values simultaneously \cite{WM_reim1, WM_reim2}.
Significantly a recent work has clarified a weak measurement is not always a good strategy in estimating a weak value:
Rather a strong measurement can be efficient to determine a weak value experimentally \cite{WM_st1, WM_st2}.
Nonetheless, for an experimental verification of a fundamental issue like observation of a quantum paradox, it would be essential to perform measurement without disturbance on the quantum system.

In the previous case of the path state of a photon, a weak measurement on a qubit system is available (i.e. which-way measurement, $|k\rangle\langle k|$ or $1-|k\rangle\langle k|$).
Actually it has been known that a CNOT operation with another qubit of a meter (measurement apparatus) achieves a weak measurement \cite{WM_qubit} as shown in figure \ref{fig:WM}(a).
The qubit to be measured (signal) in $a|0\rangle_s+b|1\rangle_s$ is correlated with the meter qubit in $\gamma|0\rangle_m+\bar\gamma|1\rangle_m$ ($\gamma\ge\bar\gamma\ge0$) by the CNOT, which results in $(a\gamma|0\rangle_s+b\bar\gamma|1\rangle_s)|0\rangle_m+(a\bar\gamma|0\rangle_s+b\gamma|1\rangle_s)|1\rangle_m$.
The correlation strength (measurement strength) is represented by $G=\gamma^2-\bar\gamma^2$ ($0\le G\le 1$).

In this setup the normalized readout, $R(1|\phi)$, corresponding to the pointer of the measurement apparatus, gives the weak value as follows, 
\begin{eqnarray}
R(1|\phi)&=&\frac{Prob(1|\phi)-\bar{\gamma}^2}{G},  \label{eq:norm} \\
&\rightarrow& {\rm Re}\langle|1\rangle\langle 1|\rangle_{\bf w} \ (G\rightarrow 0),
\end{eqnarray}
where $Prob(1|\phi)$ represents the probability of observing the meter as $|1\rangle_m$ under the success of the post-selection, $|\phi\rangle_s$.
Actually a polarization has been often used as the meter for a weak measurement on the path state of a photon \cite{W_Hardy2, W_negS, W_negW} as shown in figure \ref{fig:WM} (b).

On the other hand, the estimation of a weak value from the post-selection probability, equations (\ref{eq:post_im}) and (\ref{eq:post_re}), has an advantage that we need not prepare an ancilla system to play a role of a meter;
As mentioned later, saving the physical resource will be beneficial in a joint weak measurement.
It also achieves no disturbance in the limit of $\theta,\alpha\rightarrow 0$, albeit such an inference of a weak value is different from a weak measurement straightforwardly.

For example, according to equation (\ref{eq:post_re}), the real part of a weak value can be inferred as follows,
\begin{eqnarray}
n_{\rm est}&=&\frac{1}{2\alpha}\bigg(1-\frac{Prob(\phi)}{|\langle \phi|\psi\rangle|^2}\bigg)    \label{eq:est} \\
&\rightarrow& {\rm Re}\langle|1\rangle\langle 1|\rangle_{\bf w} \ (\alpha\rightarrow 0).
\end{eqnarray}
So as to show the above estimation is comparable to a weak measurement in practice, I calculated equations (\ref{eq:norm}) and (\ref{eq:est}) when a weak value is $\langle|k\rangle\langle k|\rangle_{\bf w}=-1$ for the path state of a photon as shown in figure \ref{fig:est} (a) and (b) respectively.
Clearly the inference of a weak value from the post-selection probability is adequate for practical use.

A weak value in the system more than 2 qubits is called a joint weak value, say $\langle|kl\rangle\langle kl|\rangle_{\bf w}$.
It is known that a joint weak value can be estimated from a correlation of pointers in the higher order on the measurement strength like $G^2$ \cite{W_Hardy1, WM_joint, WM_joint2, WM_joint3}.
Be that as it may, to perform a joint weak measurement straightforwardly, entangled meter qubits are generally needed \cite{W_Hardy2, W_pigeon} as long as considering local operations, where `local' means a signal qubit is interacted with only the corresponding meter qubit (not the other meter qubits).

However such an entangled meter is not needed in inferring a weak value from the post-selection probability;
Alternatively the c-number acts on only the corresponding term: $|kl\rangle\rightarrow C|kl\rangle$.
For example, the polarization can be free for a signal qubit in the case of a photon in figure \ref{fig:WM} (b).
Then it will practically be easy to perform an experiment involving a joint weak value by using such hybrid signals (the path and the polarization) of a single photon with experimental determination of the joint weak value.

\section{Summary}
\label{sec:sum}
The characteristic of the imaginary part of a weak value has been regarded as be different from the one of the real part, especially, in the context of a weak measurement.
However I have shown that the real and imaginary parts could appear in an equal footing as the back-action in a post-selection probability.
By observing the post-selection probability, both of the real and imaginary parts can be experimentally inferred in the limit of no disturbance on the system.
Such an estimation of a weak value has an advantage because of saving an additional system of a measurement apparatus.
Actually the back-action itself have no direct relation to weak measurement:
In my discussion just an optical component to provide a c-number has been assumed, which has no observable variables to show some result of measurement, namely, a pointer.
It relies on a weak value how the component participates in the post-selection probability.

Besides an application of estimating a weak value, I also expect that the significance of a weak value will be more clarified:
Since the real and imaginary parts could be discussed in an equal footing, the imaginary one could serve with newfound reality, as a real one has been an affinity to a probability.

\bibliography{mybibfile.bib}

\end{document}